\documentclass{aa}
\usepackage{epsfig}
\usepackage{amsmath}
\usepackage{txfonts}
\usepackage{natbib}
\usepackage{bm}

\begin{document}

\title{Can we rely on EUV emission to identify coronal waveguides?}
\authorrunning{Kohutova et al.}
\titlerunning{Identifying coronal waveguides in 3D MHD simulations}
\author{P. Kohutova\inst{1,2}, P. Antolin\inst{3}, M. Szydlarski\inst{1,2}, and N. Poirier \inst{1,2}}
\institute{Rosseland Centre for Solar Physics, University of Oslo, P.O. Box 1029, Blindern, NO-0315 Oslo, Norway\\
\email{petra.kohutova@astro.uio.no}
\and
Institute of Theoretical Astrophysics, University of Oslo, P.O. Box 1029, Blindern, NO-0315 Oslo, Norway
\and
Department of Mathematics, Physics and Electrical Engineering, Northumbria University, Newcastle Upon Tyne NE1 8ST, UK}
\date{Received; accepted}

\abstract
{Traditional models of coronal oscillations rely on modelling the coronal structures that support them as compact cylindrical waveguides. Recently, an alternative model of the structure of the corona has been proposed, where the thin strand-like coronal loops observable in the EUV emission are a result of line-of-sight integration of warps in more complex coronal structures, referred to as the coronal veil model.}
{We extend the implications of the coronal veil model of the solar corona to models of coronal oscillations.} 
{Using the convection-zone-to-corona simulations with the radiation-magnetohydrodynamics code Bifrost, we analysed the structure of the self-consistently formed simulated corona. We focus on the spatial variability of the volumetric emissivity of the Fe IX 171.073 {\AA} EUV line, and on the variability of the Alfvén speed, which captures the density and magnetic structuring of the simulated corona.  We traced features associated with large magnitudes of the Alfvén speed gradient, which are the most likely to trap MHD waves and act as coronal waveguides, and looked for the correspondence with emitting regions which appear as strand-like loops in line-of-sight-integrated EUV emission.}
{We find that the cross-sections of the waveguides bounded by large Alfvén speed gradients become less circular and more distorted with increasing height along the solar atmosphere. The waveguide filling factors corresponding to the fraction of the waveguides filled with plasma emitting in the given EUV wavelength range from 0.09 to 0.44. This suggests that we can observe only a small fraction of the waveguide. Similarly, the projected waveguide widths in the plane of the sky are several times larger than the widths of the apparent loops observable in EUV.}
{We conclude that the 'coronal veil' structure is model-independent. As a result, we find a lack of straightforward correspondence between a peak in the integrated emission profile which constitutes an apparent coronal loop and regions of plasma bound by a large Alfvén speed gradient acting as waveguides. Identifying coronal waveguides based on emission in a single EUV wavelength is not reliable in the simulated corona formed in convection-zone-to-corona models.}


\maketitle

\section{Introduction}
Coronal loops, the basic building blocks of the solar corona, are the thin strand-like structures most commonly observed in extreme-ultraviolet (EUV) and soft X-ray (SXR) emission. These correspond to the coronal plasma confined inside magnetic structures, therefore tracing the topology of the coronal magnetic field. This gives coronal loops an arch-like appearance, with typical lengths from tens to hundreds of megameters \citep{reale_2014}. These are typically few hundred km in the diameter, however, higher resolution observations with instruments such as Hi-C and the High-Resolution Imager (HRI) of the Extreme Ultraviolet Imager (EUI) of Solar Orbiter (SolO) have revealed fine-scale coronal loop structure previously unresolved by comparably lower resolution imagers such as the Atmospheric Imaging Assembly on board Solar Dynamics Observatory (SDO/AIA) \citep{peter_2013, williams_2020, Antolin_2023AA...676A.112A}. Cool tracers of the coronal magnetic field such as coronal rain \citep{antolin_2012, kohutova_2016} are also observable by high-resolution ground-based instruments, these provide further insight into the fine scale-structure of the corona \citep{scullion_2014}.

The solar corona is a dynamic environment and the coronal magnetic field changes and evolves continuously. Recent observational evidence suggests complex density structuring \citep{berghmans_2023}. Simplified models of static coronal loops neglect this spatial and temporal variability of the solar atmosphere. In order to account for realistic magnetic field configurations and density structuring in the solar corona, a more self-consistent approach to modelling the evolution of coronal structures is necessary. One such approach involves using self-consistent convection-zone-to-corona simulations, where the structuring and the evolution of the corona is driven entirely by the dynamics of the lower solar atmosphere. It can be therefore argued that the simulated corona formed in such models is more realistic. 

The structure of the corona in such models has been analysed by \citet{malanushenko_2022}, where the relationship between integrated synthetic EUV emission and the three-dimensional structure of the volumetric emissivity has been investigated using a self-consistent MURaM simulation. Contrary to the popular image of coronal loops as well-defined plasma cylinders clearly distinct from the surroundings, the emitting structures seen in this type of models are much more complex, and the strand-like appearance of coronal loops visible in the synthetic EUV emission comes from wrinkles in such shapes integrated along the line-of-sight akin to folds in a veil. This 'coronal veil' coronal structure is present in other convection-zone-to-corona models, including Bifrost models \citep{gudiksen_2011}, and hence appears to be model-independent. It is worth noting that even simple 3D MHD coronal models of transverse wave propagation along flux tubes that generate shear-flow instabilities (such as Kelvin-Helmholtz, KH) can produce strand-like structures due to the compressive physical processes of the KH roll ups \citep{Antolin_2014ApJ...787L..22A}. Hence, the fine-scale structure of the corona can be independent of the motions of the footpoints of flux tubes at photospheric or chromospheric heights.

\begin{figure*}
	\includegraphics[width=43pc]{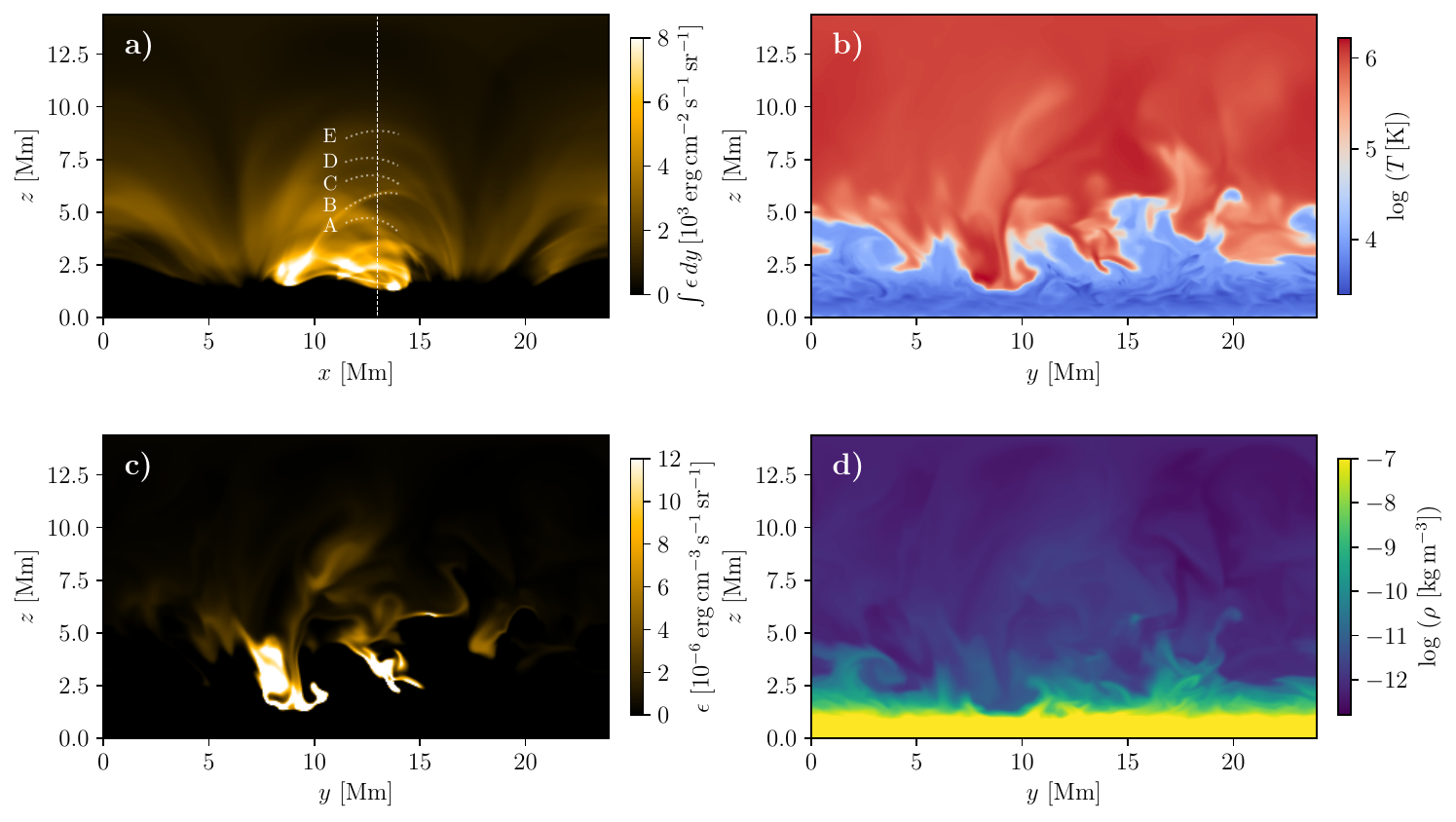}
	\caption{Coronal structure in the simulation. We show the LOS-integrated Fe IX emission intensity in the $x-z$ plane (a). The dashed line at x = 13 Mm marks the y-z plane along which the subsequent slices are taken. The dotted lines mark a portion of the loop axis for loops A - E. Vertical slices taken across the coronal loops at $x=13$ Mm showing the plasma temperature (b), volumetric emissivity (c) and the plasma density (d).}
	\label{fig:context_full}
\end{figure*}

\begin{figure*}
	\includegraphics[width=43pc]{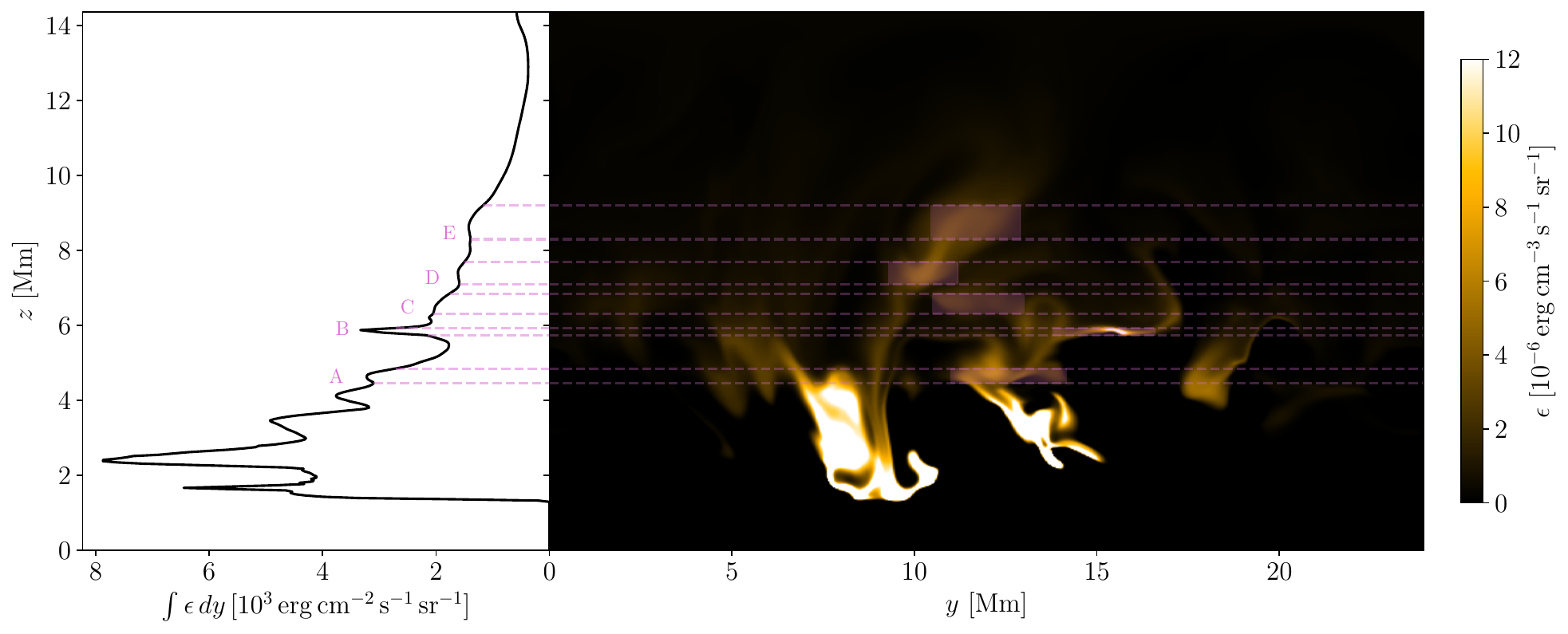}
	\caption{Vertical slice across the volumetric emissivity at $x=13$ Mm (right), and the corresponding $x=13$ Mm LOS-integrated emission profile (left). The local emission peaks corresponding to the loops A-E are marked by purple dashed lines. These also mark the z-coordinates of the contributing regions in the emissivity slice. The dominant emitting regions corresponding to each loop are marked by purple rectangles.}
	\label{fig:e_profile}
\end{figure*}

In the case of oscillations in coronal structures, the coronal loops act as waveguides, in which the inhomogeneity of physical properties provides a boundary, often leading to an onset of standing modes in the structures. The waveguide boundaries are determined by the transverse variation in the density at edges of the loop, or more generally, by the variation of the Alfvén speed \citep{nakariakov_1996}.  

The classical model for coronal oscillations relies on the magnetised cylinder model \citep{edwin_1983} and assumes the individual coronal loops have approximately circular cross-sections and are to a large degree decoupled from the surrounding background plasma. Such model forms the basis for coronal seismology, a widely used method which employs coronal oscillation parameters for diagnostics of physical properties of the coronal plasma, which are otherwise difficult to measure directly, such as coronal densities and magnetic fields \citep{nakariakov_2020}. The complexity of the coronal structure seen in self-consistent models of the solar atmosphere, however, clearly contradicts such a simplified picture. The impact of such structure on the accuracy of coronal seismological methods largely based on the magnetic cylinder approximation for coronal loops is still not clear.

In this paper we analyse the structure of the corona in a three-dimensional convection-zone-to-corona simulation using the radiation-MHD code Bifrost. We aim to identify the waveguides in the coronal veil based on the variation of Alfvén speed in three dimensions, and match those to the strand-like features appearing in the forward-modelled EUV emisssion. This tells us how valid our approximations about the 3-dimensional structure of oscillating coronal loops are.

\begin{figure*}
	\includegraphics[width=43pc]{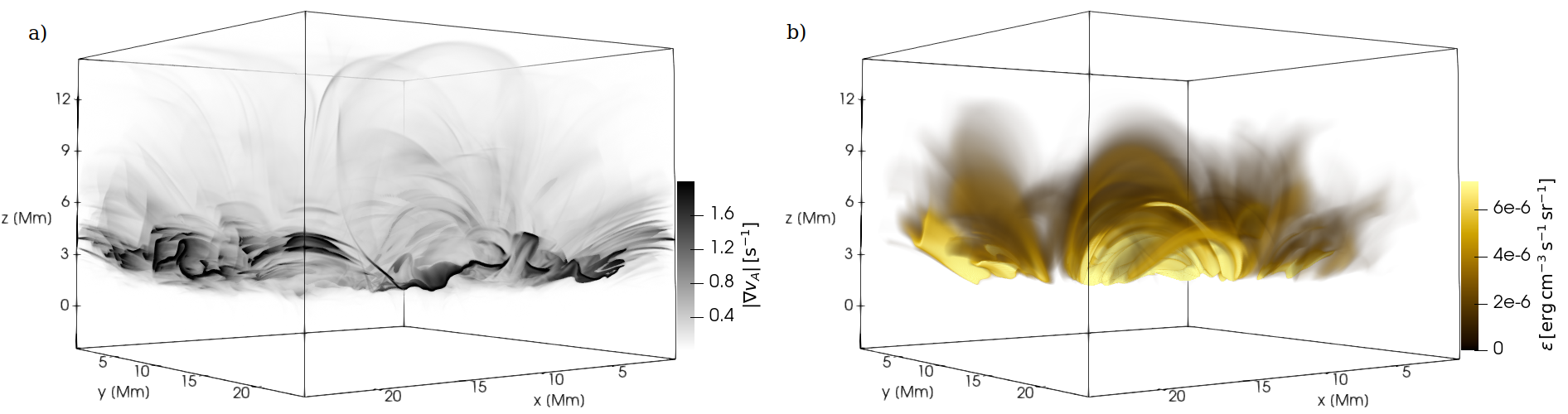}
	\caption{Three dimensional rendering of $|\nabla v_{\mathrm{A}}|$ (a) and the volumetric emissivity $\epsilon_{\mathrm{Fe}\,\mathrm{IX}}$ (b). The animation is available online.}
	\label{fig:3d_still}
\end{figure*}

\begin{figure*}
	\includegraphics[width=43pc]{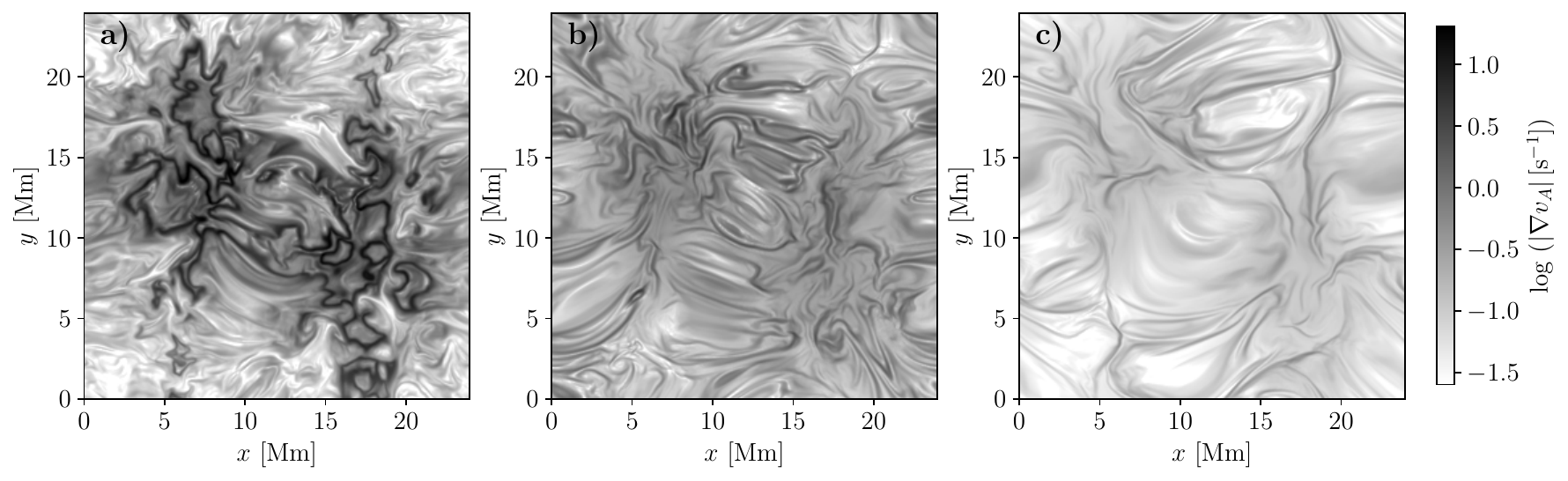}
	\caption{Horizontal slices across $|\nabla v_{\mathrm{A}}|$ at $z=2$ Mm (a) $z=5$ Mm (b) and $z=10$ Mm (c). The animation is available online.}
	\label{fig:vg_variation}
\end{figure*}

\section{Numerical model} 
\label{section:model}
We analyse the structure of the corona formed in the numerical simulation of a magnetically enhanced network spanning from the upper convection zone to into the coronal heights using the radiation-MHD code Bifrost \citep{gudiksen_2011}. Bifrost is a 3D radiation-MHD code that solves resistive MHD equations on a staggered cartesian grid. Built on top of the Stagger code \citep{stein_2024}, it includes radiative transfer with scattering in the photosphere and lower chromosphere, and parameterized radiative losses and heating in the upper chromosphere, transition region, and corona. The simulation used in this work also accounts for field-aligned thermal conduction and non-equilibrium ionization of hydrogen in the equation of state. The size of the numerical box is 24 $\times$ 24 $\times$ 16.8 Mm spanning from 2.4 Mm below the photosphere to 14.4 Mm in the corona, and the grid resolution is 504 $\times$ 504 $\times$ 496. The simulation boundary conditions are periodic in the $x$ and $y$ directions and open in the $z$ direction. The top boundary uses characteristic boundary conditions, designed to transmit disturbances with minimum reflection \citep{gudiksen_2011}. The flows are allowed to pass through the bottom boundary and the magnetic field is passively advected - no additional magnetic field is introduced into the domain. The numerical setup is described in more detail in \citet{kohutova_2023}. The analysed snapshot corresponds to the $t = 980$ s timestep of the extended simulation run described therein. In such a model, the coronal evolution and heating is driven by the dynamics of the lower solar atmosphere, and the associated convective motions. The high coronal temperatures are maintained primarily through the Joule and viscous heating associated with magnetic braiding, the heating in the vast majority of the domain is hence 'self-consistent'. The temporal variability of the simulated corona is significantly more complex compared to more idealised models, as the footpoints of the magnetic structures are dragged around by the convective motions. Similarly, the coronal structure is driven by the dynamics of the lower solar atmosphere; the magnetic configuration is initialised from two opposite polarity patches, these are quickly swept into the intergranular lanes by the convective downflows. The structure of the corona is therefore free from making a priori assumptions about the shape of the coronal loops. There are several coronal structures with densities higher than the surrounding plasma, these are filled by the chromospheric evaporation in response to heating events \citep[e.g.][]{Kohutova_2020}. 

To analyse the appearance of the coronal loops in the model in EUV, we calculate the synthetic emission in the Fe IX 171.073 {\AA} coronal line. The emission intensity $I_{\lambda}$ for the optically thin coronal EUV lines corresponds to the integral of the volumetric emissivity at the specific wavelength along the line-of-sight:
\begin{equation}
    I_{\lambda} = \int \phi_{\lambda} \epsilon_{\lambda 0} dl   
\end{equation}
where $\phi_{\lambda}$ is the gaussian line profile accounting for the Doppler shift due to the velocity along the line-of-sight (LOS) and for the thermal line broadening, and $\epsilon_{\lambda 0}$ is the volumetric emissivity of the given spectral line at the rest wavelength:
\begin{equation}
    \epsilon_{\lambda 0} = n_e^2 G_{\lambda 0}(n_e, T)
\end{equation}
Here $n_e$ is the electron density, $T$ is the plasma temperature and $G_{\lambda 0}$ is the contribution function of the specific spectral line calculated using the CHIANTI atomic database v.10 \citep{zanna_2021}. We calculate $G_{\lambda 0}$ values of the Fe IX 171.073 {\AA} coronal line on a $200 \times 3000$ grid of temperatures and densities to create a look-up table for speeding up the calculation. We do this for the temperature range from $\log T = 4.0$ to $\log T = 7.0$ and for the density range $\log n_{\mathrm{e}} = 8$ to  $\log n_{\mathrm{e}} = 11$ (in cgs units). We then use cubic spline interpolation in $T$ and zero-order interpolation in $n_{\mathrm{e}}$ to determine the value of $G_{\mathrm{Fe IX}}$ and subsequently $\epsilon_{\lambda 0}$ at each simulation grid point. As we are interested in the three dimensional structure of the coronal EUV emission, $\epsilon_{\mathrm{Fe IX}}$ is the main quantity we will focus on in the analysis below.

\begin{figure}
	\includegraphics[width=21pc]{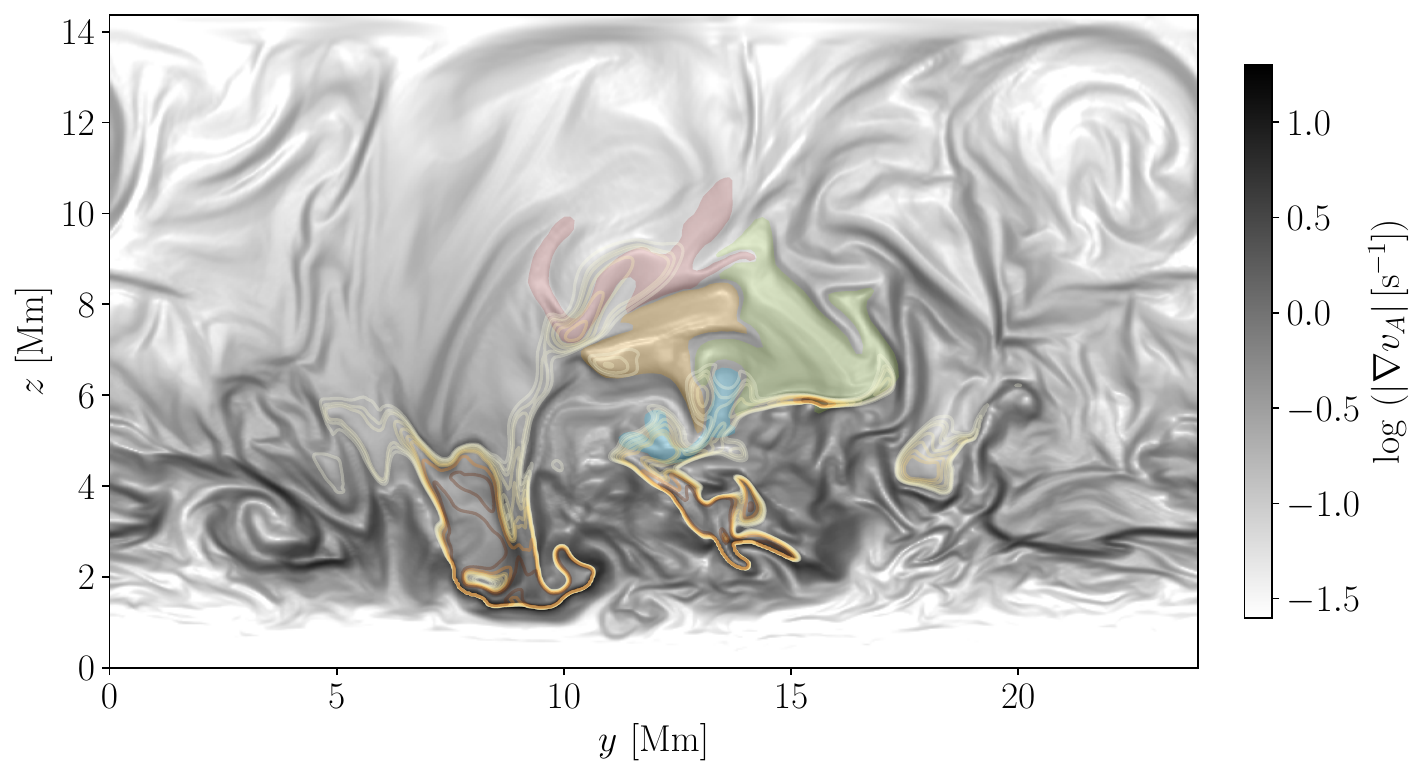}
	\caption{The $x=13$ Mm slice across $|\nabla v_{\mathrm{A}}|$. The cross-sections of waveguides encompassing dominant emitting regions for each loop are shown in blue (waveguide 1), green (waveguide 2), orange (waveguide 3) and red (waveguide 4). The emissivity contours are overplotted.}
	\label{fig:va_grad}
\end{figure}

\begin{figure*}
	\includegraphics[width=43pc]{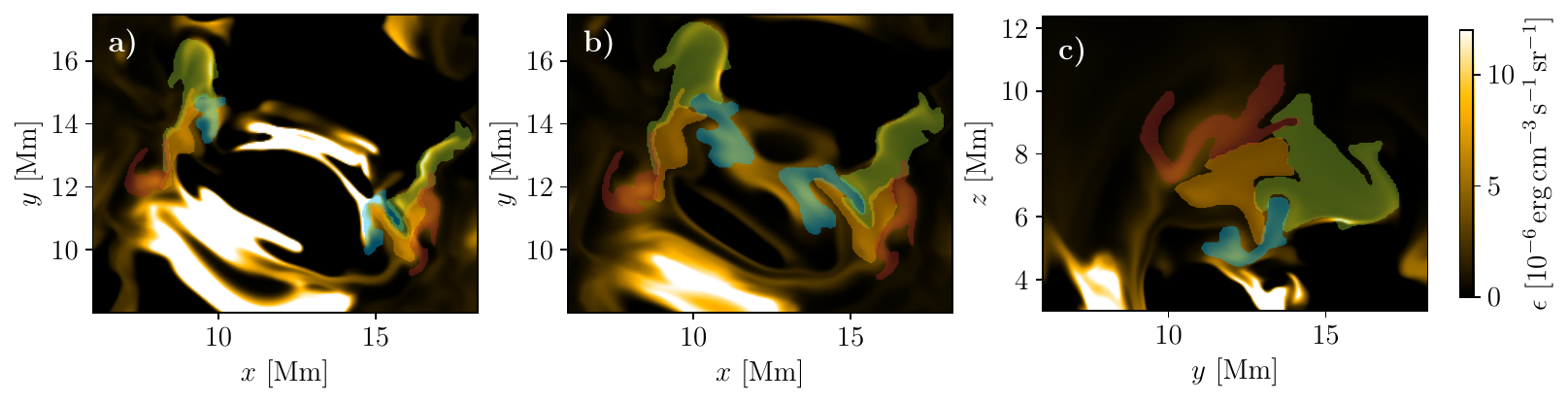}
	\caption{Cuts across the volumetric emissivity along the $z=3$ Mm (a), $z=4.5$ Mm (b), and $x=13$ Mm (c) plane corresponding to cuts across the coronal loop footpoints, coronal loop legs and the coronal loop apex. The corresponding cuts across the individual waveguides are overplotted in blue (waveguide 1), green (waveguide 2), orange (waveguide 3) and red (waveguide 4).}
	\label{fig:wg_variation}
\end{figure*}

\section{Coronal waveguides}
\label{section:waveguides}
We use the 3-dimensional volumetric Fe IX emissivity to examine the coronal structures in the simulation. Vertical slices across the simulation domain at $x = 13$ Mm showing the structure of the plasma density, temperature and the corresponding volumetric emissivity, along with the line-of-sight (LOS) integrated Fe IX emission in the $x-z$ plane are shown in Fig. (\ref{fig:context_full}). The synthetic LOS-integrated emission contains several coronal loops, which are aligned with the magnetic field connecting the two magnetic polarity patches at $z = 0$ Mm. We partially mark the axis outline of 5 distinct loops and label them A-E. The width of the coronal loops varies from $\sim$ 1 Mm for the more diffuse loops (e.g. loop E) down to 200 km for the most fine-scale strands (loop B). The coronal structure visible in the $x=13$ Mm slice across the coronal loops shows complex wrinkled surfaces rather than containing clearly defined loop cross-sections. The transverse structuring of the coronal loops is consistent with the 'coronal veil' interpretation discussed by \citet{malanushenko_2022}. The temperature and density structure in the identical slices follows the complex shapes seen in the emissivity slice.

To cross-correlate the loops visible in the LOS-integrated emission with the emitting features in the volumetric emissivity, we examine the emission profile along the dashed line intersecting the loops visible in the synthetic EUV emission at the apex (Fig. \ref{fig:e_profile}). The local peaks in the emission profile are matched to the loops marked in Fig. \ref{fig:context_full}. The $z$-coordinates of the loop boundaries are then plotted over the $x=13$ Mm emissivity slice to mark out the emitting features contributing to peaks A-E when integrated along the $y$-axis. We find that multiple coronal structures with varying $y$-coordinates contribute to a single peak in the emission profile, which we identify as a single loop, due to the LOS superposition. We note that the features located at lower coronal heights have a very high degree of LOS superposition, which subsequently decreases at greater coronal heights. We highlight the dominant contributing emission regions corresponding to each loop. The dominant features in the EUV emissivity have irregular shapes, the cross-sections of the corresponding coronal structures therefore cannot be approximated as circular. 

We show the full three-dimensional rendering of the volumetric emissivity in Fig. \ref{fig:3d_still} and the associated animation, which demonstrates the line-of-sight effects when viewing the entire simulation domain from different directions. Several features with enhanced emissivity aligned with the coronal magnetic fields are clearly visible in the simulation domain, with varying thickness and complex cross-sections. In particular, we find that the primary contributing sources to features with strand-like appearance in the LOS-integrated EUV emission in the $x-z$ plane are thin sheet-like structures with elongated spatial extent in the $y$-direction (this is most prominent for the loops A, B and C).

We further determine the variation in the Alfvén speed $v_{\mathrm{A}} = B/\sqrt{\mu_0 \rho}$ in the simulated corona by calculating the Alfvén speed gradient $\nabla v_{\mathrm{A}}$, which highlights the regions with inhomogeneities in the physical properties of the coronal plasma acting as waveguide boundaries (Fig. \ref{fig:3d_still}). The spatial distribution of $|\nabla v_{\mathrm{A}}|$ across the simulation domain has a similar `veil-like' structure as the EUV emissivity, and the regions with the strongest $|\nabla v_{\mathrm{A}}|$ form coherent surfaces extending across several Mm, suggesting both large spatial extent and complex shapes of the regions in the corona acting as individual waveguides. Fig. \ref{fig:vg_variation} shows the variation of the structures corresponding to enhanced $|\nabla v_{\mathrm{A}}|$ with height. In the figure, we show the horizontal slices of $|\nabla v_{\mathrm{A}}|$ at three different heights at $z = 2$ Mm corresponding to the loop footpoints close to the transition region, at $z = 5$ Mm corresponding to the loop legs in the lower corona and at $z = 10$ Mm at coronal heights corresponding to the location of the loop apex (this is only approximate, as the loop tops have significant spatial extent, see fig. \ref{fig:3d_still}). The associated animation shows the transition through the entire solar atmosphere starting from the photosphere at $z = 0$ Mm to the upper boundary of the simulation domain. This highlights how the footpoints of the structures defined by enhanced $|\nabla v_{\mathrm{A}}|$ transition from well-defined, albeit irregular closed cross-sections at $z = 2$ Mm in the transition region into the veil-like structure at coronal heights at $z = 10$ Mm.

In order to determine the characteristics of the individual waveguides which are associated with the coronal loops A-E, or rather with the source regions responsible for the appearance of the loops, we plot the $x= 13$ Mm slice through $|\nabla v_{\mathrm{A}}|$ (Fig. \ref{fig:va_grad}). We trace the edges visible in the $|\nabla v_{\mathrm{A}}|$ plot while identifying enclosed shapes that form the waveguide cross-section in the $x= 13$ slice overlapping the loop emission source regions marked in Fig. \ref{fig:e_profile} (also shown by emissivity contours in Fig. \ref{fig:va_grad}). We then trace the magnetic connectivity of the waveguide boundaries to obtain the full three-dimensional structure of each waveguide. We terminate the integration of $\vec{B}$ at $z = 1.5$ Mm, as below this height there is no emission in the coronal lines. We note that the $|\nabla v_{\mathrm{A}}|$ varies along the waveguide boundaries (albeit not as strongly as across the waveguide boundaries), making separation difficult in places with weaker gradients. 

\begin{figure*}
	\includegraphics[width=43pc]{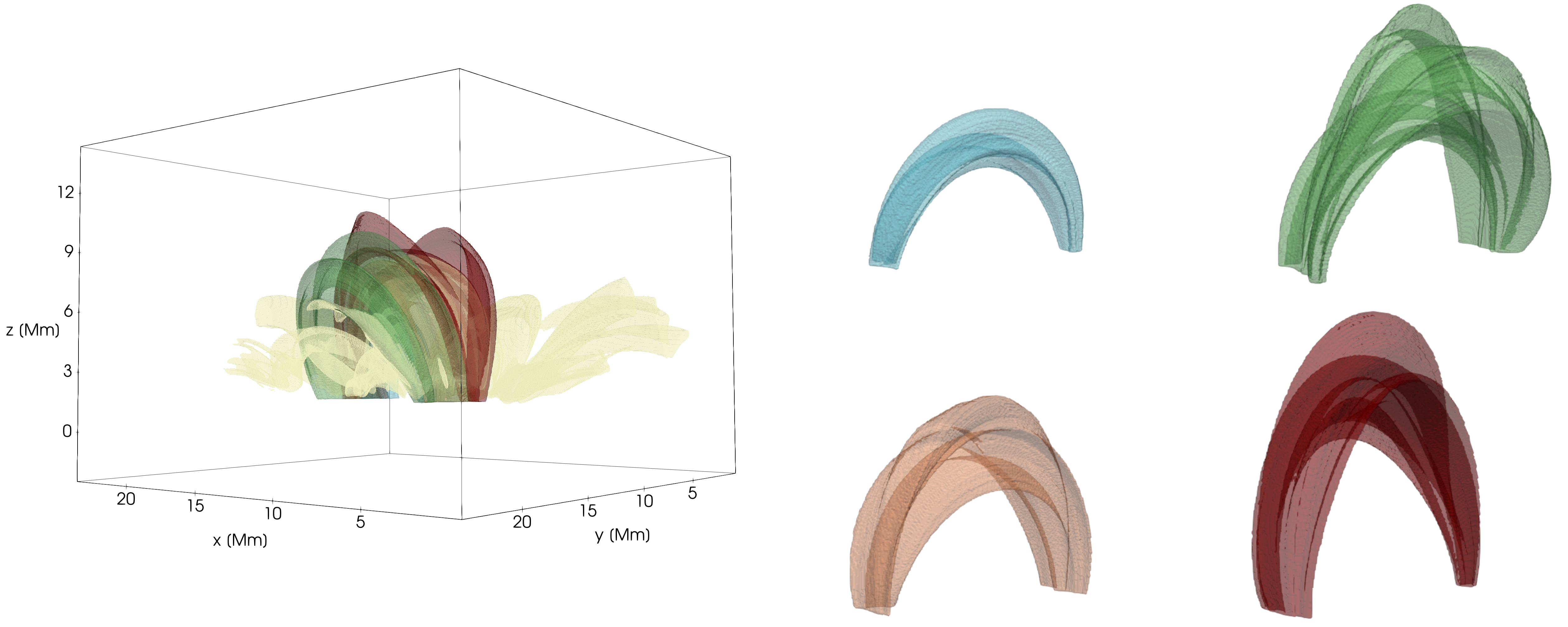}
	\caption{The 3-dimensional structure of the waveguide 1 (blue), waveguide 2 (green), waveguide 3 (orange) and waveguide 4 (red). The volumetric emissivity above the threshold is shown in yellow. An animation of this figure is available online.}
	\label{fig:waveguides}
\end{figure*}

Fig. \ref{fig:wg_variation} shows cuts across the volumetric emissivity and the corresponding waveguide cross-sections along different planes corresponding to the loop footpoints close to the transition region, loop legs in the lower corona and the coronal loop apex. The full three-dimensional structure of the waveguides is shown in Fig. \ref{fig:waveguides}. Even though some of the boundaries of the strong emissivity regions and strong Alfvén speed gradients are aligned (Fig. \ref{fig:va_grad}), we find that the waveguides generally encompass regions larger than the emitting coronal structures. This is particularly true to the emitting structures appearing as loops D and E in the line-of-sight integrated emission, these 2 structures are both part of a much larger waveguide. 

The boundary surfaces of the waveguides are complex and contain multiple folds aligned with the direction of the magnetic field strength. We quantify the complexity of the surface of each individual waveguide by calculating a circularity index $I_\mathrm{c}$ at different positions in the simulation domain: in the $x=13$ Mm plane close to the apex of the magnetic loops in the simulation domain, in the $z = 3$ Mm plane close to the loop footpoints in the transition region and in the $z = 4.5$ Mm plane intersecting the loop legs in the lower corona (the waveguide cross-sections are shown in Fig. \ref{fig:wg_variation}). The circularity index is defined as $I_\mathrm{c} = 4 \pi A/d^2$, where $A$ is the cross-sectional area of the waveguide and $d$ is the circumference of the waveguide cross-section. It is a measure of a departure of the waveguide cross-section from a circular cross-section ($I_\mathrm{c} = 1$ for a perfect circle). The $I_\mathrm{c}$ values of both footpoints and the apex for each waveguide are listed in Table \ref{tab:properties} and vary from 0.11 to 0.41. We find that $I_c$ in general decreases with height except for the waveguide 3, being highest at the footpoints, and decreasing at the apex. There is a varying degree of asymmetry present between the left and right footpoints.

We further estimate the projected width of the waveguides, that is the total extent of each waveguide along the $z$-axis at $x=13$ Mm, when considering the LOS-projection (Table \ref{tab:properties}). This enables us to compare the waveguide extent to the apparent width of the coronal loops observed in LOS-integrated emission. These vary from 2.21 Mm to 4.42 Mm, as opposed to apparent loop widths in emission which are of the order of several hundred km (the exact values are instrument/bandpass dependent), meaning large parts of the waveguides are unaccounted for when relying on EUV observations only.

We finally calculate the waveguide filling factor, that is, the fraction of the total waveguide volume which is filled by the emitting coronal plasma. To estimate this we set $2.88\times 10^{-6}$ erg cm$^{-3}$s$^{-1}$sr$^{-1}$ as the emissivity threshold, the exact value of which is instrument-dependent (in our model the typical emissivity values for the background, quiet corona are around $1-2 \times 10^{-6}$ erg cm$^{-3}$s$^{-1}$sr$^{-1}$). Plasma with the volumetric Fe IX emissivity larger than the threshold is considered emitting and shown in yellow in Fig. \ref{fig:waveguides}. The filling factor values for the waveguides 1-4 are shown in Table \ref{tab:properties} and range from 0.09 to 0.44. 

\begin{table*}[t]
	\caption{Waveguide properties}
	\label{tab:properties}
	\centering
	\tabcolsep=0.11cm
	\begin{tabular}{ccccccccc}
		\hline\hline
		Loop & Waveguide & Filling factor & $I^{\mathrm{3 Mm}}_{\mathrm{cL}}$ & $I^{\mathrm{3 Mm}}_{\mathrm{cR}}$ & $I^{\mathrm{4.5 Mm}}_{\mathrm{cL}}$ & $I^{\mathrm{4.5 Mm}}_{\mathrm{cR}}$ & $I_{\mathrm{cA}}$ & Projected width [Mm]\\ 
		  \hline
		$\mathrm{A}$    & 1 & 0.44 & 0.41 & 0.23 & 0.26 & 0.25 & 0.27 & 2.21   \\
		$\mathrm{B}$    & 2 & 0.09 & 0.20 & 0.18 & 0.19 & 0.20 & 0.19 & 4.42  \\
		$\mathrm{C}$    & 3 & 0.12 & 0.24 & 0.22 & 0.25 & 0.22 & 0.34 & 3.43  \\
		$\mathrm{D+E}$  & 4 & 0.30 & 0.18 & 0.15 & 0.14 & 0.15 & 0.11 & 3.77  \\
		
		\hline
	\end{tabular}
	\tablefoot{Summary of waveguide properties: The waveguide filling factor, the circularity index for the left and right footpoints $I_\mathrm{cL}$, $I_\mathrm{cR}$ at $z = 3$ Mm and $z = 4.5$ Mm, the circularity index at the apex $I_\mathrm{cA}$ and the projected width at the apex.}
\end{table*}

\section{Discussion}

\subsection{The coronal structure inferred from observations vs. self-consistent MHD models}
The convection-zone-to-corona models of the solar atmosphere seem to agree on reproducing the veil-like structure of the solar corona and the absence of well-defined cylindrical coronal loops, as also seen in \citet{malanushenko_2022}. 

Even when considering MHD models initiated with simple cylindrical loop configurations, it has been found that the turbulent evolution of oscillating loops quickly leads to distortion of the loop boundaries \citep{magyar_2018, karampelas_2019}, highlighting that the models of coronal loops as long-lived confined structures might be too idealised in a real coronal environment. In a realistic corona which is highly dynamic and the coronal structures are subject to translational motion, oscillations in different modes and polarisations, and external perturbations leading to displacements of the magnetic structures \citep[e.g.][]{kohutova_2021, kohutova_2023}, it is expected that these will all affect the morphology and lifetimes of the coronal loops.

Additional complexities, such as non-cospatiality of enhanced temperature and density structures, both of which lead to increased volumetric emissivity have been highlighted in 3D MHD simulations by \citet{peter_2012, chen_2014}. \citet{peter_2012} also highlight a case where a fraction of a simulated coronal loop is subject to temperature variations large enough to lay outside of the contribution function of the synthesised emission line. This coronal loop part hence appears dark in the synthetic observations and can therefore lead to erroneous conclusions about the width, transverse density profile and the vertical cross-section variation of the coronal loop in question.

The observational analysis of the real cross-sectional structure of coronal loops remains limited. This is mainly due to the requirement of having multiple vantage points as well as sufficient spatial resolution to enable stereoscopic reconstruction, as direct probing of the three-dimensional coronal structure is not possible. This has been attempted by \citet{mccarthy_2021} using a combination of SDO/AIA and STEREO-EUVI \citep{kaiser_2008} observations, finding a lack of correlation between loop diameters seen from multiple perspectives, suggesting features traditionally identified as monolithic and cylindrical coronal loops might be in fact more complex structures.

Using an alternative approach, \citet{klimchuk_2020} used Hi-C data to analyse the cross-sectional shape of the coronal loops by investigating the relation between the coronal loop width and the emission intensity seen in the EUV emission band centred on 193 Å. A lack of correlation between coronal loop width and intensity was used as a proof of circularity of the loop cross-section, under the key assumption that there is non-negligible magnetic twist in each coronal loop. We note that anti-correlation between loop widths and intensity has been found by \citet{mccarthy_2021}.

The discussion about whether coronal loops are being better represented by confined strands or extended veils is nuanced, and very likely both scenarios occur at some point in the corona. A notable example is a recently published analysis of the polarisation of a decayless kink oscillation using multi-viewpoint data from SDO/AIA and SolO/EUI \citep{zhong_2023}, in which case the thin studied loop is well isolated from the surroundings and identifiable in both datasets. However, conclusive observational evidence pointing to either form of the coronal structure being more prevalent is still missing. Combining newly available data from SolO and SDO observations might shed more light into the question, thanks to their potential for multi-viewpoint observations and the boost in the available resolution compared to stereoscopic studies using the pair of STEREO spacecraft. However, because of the stereoscopic ambiguity discussed by \citet{malanushenko_2022}, the use of additional diagnostics might be necessary.

\subsection{The concept of a waveguide}

In the context of coronal oscillations, what is typically understood under the term 'waveguide' or 'wave cavity', is a coherent structure in the solar corona capable of trapping MHD waves and oscillating as a whole. It can guide propagating waves along and can undergo resonant, or standing mode oscillations, where the oscillation periods supported are one of the natural modes of the waveguide \citep[e.g.][]{edwin_1982, edwin_1983}, provided there is a sufficient change of physical conditions inside and outside of the waveguide, causing wave refraction and acting as a waveguide boundary \citep{nakariakov_1996}. In standard models of coronal oscillations, such boundary is typically provided by an increased density inside the loop, but uniform density with, instead, a varying magnetic field strength, is also possible \citep{Howson_2019AA...631A.105H}. In this work, we have based the waveguide identification on the regions with large values of the magnitude of the Alfvén speed gradient which act as a boundary for the magnetoacoustic waves. However, the exact behaviour of the waves at such a boundary is dependent on the wave mode and frequency. Furthermore, we note that in a realistic coronal setup, both gravitational stratification and the magnetic field expansion cause the density and the magnetic field strength to vary along the magnetic field lines leading to the variation of the Alfvén speed gradient along the magnetic field lines. 

Our analysis suggests that in the self-consistent MHD models where the corona is driven by the dynamics of the lower solar atmosphere the coronal waveguides are far from the idealised image of long and thin cylinders and such approximations are therefore not valid in these models. In order to have a clear understanding of the coronal oscillations in such a setup, it is necessary to model oscillatory behaviour of the entire waveguide, including oscillation modes and their spatial characteristics and the oscillation polarisation. We further find that waveguide boundaries follow the veil-like features of the coronal structure, leading to a `wrinkled' waveguide surface. The waveguide cross-sections are far from circular, and the circularity index of the cross-sections decreases with the increasing height in the solar atmosphere. This can be potentially caused by more complex dynamics higher in the simulated atmosphere, where small perturbations at the chromospheric heights lead to larger disturbances in the corona because of the magnetic field expansion and density stratification.

We note that the emissivity structures are likely to vary on the timescales given by the sound speed, as they are density and temperature dependent whereas the spatial structure of the waveguides will vary on the Alfvén timescales. The mismatch between the two timescales adds another layer of complexity to the problem, and a detailed analysis of the temporal evolution of both structures is necessary to quantify this. The evolution of coronal oscillations and the temporal variation of the veil emissivity structure will be addressed in a follow-up work.

Our analysis highlights the complexity and a lack of clear 1-1 correspondence between a peak in the integrated emission profile which constitutes a loop cross-section, the regions of plasma along the LOS contributing to said emission, and the regions of plasma bound by a large Alfvén speed gradient. This is well illustrated by the waveguide 4 where what appears as two distinct loops D and E are in fact both a part of a single waveguide with large spatial extent.

Another factor that has to be considered when discussing waveguide characteristics is the waveguide skin depth. This corresponds to the exponential decay length at which the oscillation is evanescent at the boundary of the waveguide. Generally speaking, the skin depth is also mode and frequency dependent, in the idealised scenario of a discontinuous boundary between a waveguide and the ambient plasma it is given by \citep{hindman_2021}:

\begin{equation}
    \frac{1}{\Delta^2} = k_{z}^2 - \frac{\omega^2}{v_{\mathrm{e}}^2}
\end{equation}

 $\Delta$ here corresponds to a length scale to which the oscillations extend beyond the waveguide, and is independent from the transverse geometry, $\omega$ is the frequency, $k_{z}$ is the longitudinal wavenumber and $v_{\mathrm{e}}$ is the Alfvén speed in the ambient plasma. We note this is different from the 'skin depth' in some studies of coronal loop oscillations, where the term is simply used to refer to the width of the boundary layer in an idealised model of a cylindrical flux-tube, where the density linearly changes from the internal (the coronal loop density) to the external value (ambient plasma density). This quantity is usually defined in the model setup a priori and is not directly related to the actual spatial extent of the wave field into the ambient plasma surrounding the waveguide. 

The effect of the finite skin depth is that different waveguides located close to an oscillating waveguide will be affected if their separation is closer than the skin-depth of that particular mode. This is the case in the coronal structures analysed in this work, as the waveguides 1-4 lie very close to each other. The coupling of modes of an oscillating coronal loop to the modes of the surrounding arcade has been investigated by\citet{hindman_2021}, highlighting the possibility of an arcade resonance appearing as resonant modes of individual loops and potentially leading to errors in the seismological estimates. Despite the model being limited to 2D, a very similar coupling can be expected in three dimensions. Similarly, \citet{Luna_2019AA...629A..20L} has investigated oscillation modes of a cluster of strands, finding a large degree of collective behaviour.

The observational evidence seems to also suggest that coronal loops in an active region do not oscillate in isolation, but are often coupled to oscillations of surrounding structures \citep{verwichte_2004, verwichte_2009, jain_2015, tian_2016, li_2023}. Due to the nature of the EUV observations, however, only the evolution of the emitting loops can be analysed.

This suggests the connection of the coronal waveguides to the ambient plasma should be also considered, as opposed to modelling coronal loops as isolated entities. The downside is of course the complexity of modelling the collective oscillation of the whole active region loop system. This is not to say that well-defined coronal loops evolving independently of their surroundings do not exist. For instance, loops that catastrophically cool and exhibit coronal rain, considered to be in a state of thermal non-equilibrium and thermal instability \citep[known as `TNE-TI' scenario][]{Antolin_Froment_2022}, can be considered as isolated coronal waveguides, and \citet{Sahin_2022ApJ...931L..27S} show that they can be prevalent over an active region.

In the above analysis we have also estimated the waveguide filling factors, that is a fraction of the coronal waveguide which is filled with plasma emitting in EUV. We note that the exact values will depend on the resolution and the sensitivity of the instrument, as well as on the wavelength extent of the instrument bandpass. The overarching conclusion which is still valid regardless of the above is that we can only observe a small part of the actual coronal waveguide. The discussion about filling factors in the corona is usually focused on the volume filling factors of multistranded coronal loops, to quantify the number of strands emitting in the given channel \citep{peter_2013}, where the outer envelope of the loop is determined from EUV observations. In our work, we focus purely on the relationship between the EUV volumetric emissivity and regions bounded by large Alfvén speed gradients.

The natural next step is to investigate the link between the temporal evolution and oscillations in the EUV emission and the actual dynamics of the oscillating plasma structures in 3 dimensions; this will be done in a follow-up study. The impact of such non-ideal waveguides on the accuracy of the coronal seismology will be also addressed. 

\subsection{Model limitations}
One of the obvious limitations of the model used in this study (and by extension, of most purely MHD codes used for simulating the solar atmosphere with high realism) is the artificially low Reynolds number. This affects the formation of the fine-scale structure in the model due to the code being inherently diffusive. Despite this, the observed fine-scale structure of the corona, as appears in LOS-integrated emission, is still reproduced \citep{malanushenko_2022}. The fine-scale structure is additionally affected by the spatial resolution of the model.

A possibility of course exists that the coronal structure reproduced in convection-zone-to-corona simulations is simply an artifact of the model. However, the fact that the same veil-structure has been reported in self-consistent MURaM simulations suggests such structure of the corona is model-independent. Furthermore, even models that only include a coronal waveguide will also be subject to strong LOS superposition that leads to apparent strand formation \citep{Antolin_2014ApJ...787L..22A}. The properties of coronal oscillations reproduced in such models such as the excitation of different modes, harmonics and the oscillation damping timescales analysed so far agree with the observations \citep{kohutova_2021, kohutova_2023}. More generally, a number of complex phenomena has been successfully reproduced by the convection-zone-to-corona models including solar flares \citep{cheung_2019}, surges \citep{nobrega_2016}, the formation of coronal rain \citep{Kohutova_2020}, and coronal brightpoints \citep{nobrega_2023}, all suggesting a high degree of realism of the solar atmosphere formed in such models.

Regardless of how accurate the solar corona formed in the convection-zone-to-corona models really is, this work highlights the possibility of misinterpretation of observations of oscillations of coronal structures, namely of what actually constitutes a waveguide. This can occur as a result of the ambiguities induced by the LOS integration combined with the finite temperature range sensitivity of a given bandpass and the fact that both the density and the magnetic structure govern the propagation, and potentially the trapping of MHD waves in the corona.

\section{Conclusions}
 We have extended the implications of the `coronal veil' model of the solar corona to models of coronal oscillations. Using the convection-zone-to-corona simulations with the radiation-MHD code Bifrost, we analysed the structure of the simulated corona self-consistently formed in such a model. We conclude that the `coronal veil' structure is model-independent and appears in convection-zone-to-corona models other than MURaM.
 
We focused on the spatial variability of the volumetric emissivity of the Fe IX 171.073 {\AA} EUV line, and on the variability of the Alfvén speed, which captures the density and magnetic structuring of the simulated corona. We traced features boundaries of which are associated with large magnitudes of the Alfvén speed gradient and which are the most likely to trap MHD waves and act as coronal waveguides, and looked for the correspondence with emitting regions which appear as strand-like loops in LOS-integrated EUV emission.

We have found that the cross-sections of the waveguides bounded by large Alfvén speed gradients become less circular and more distorted with increasing height along the solar atmosphere. Small filling factors corresponding to the fraction of the waveguides filled with plasma emitting in the given EUV wavelength suggest that we can observe only a small fraction of the waveguide. Similarly, the projected waveguide widths in the plane of the sky are several times larger than the widths of the apparent loops observable in EUV. Our results point to a lack of straightforward correspondence between a peak in the integrated emission profile which constitutes an apparent coronal loop and regions of plasma bound by a large Alfvén speed gradient acting as waveguides. This may lead to incorrect assumptions about the size, shape and the transverse density structuring of the oscillating coronal features. Identifying coronal waveguides based on emission in a single EUV wavelength is therefore not reliable in the simulated corona formed in convection-zone-to-corona models.

\label{section:conclusions}

\begin{acknowledgements}
PK and NP acknowledge funding from the Research Council of Norway, project no. 324523. This research was also supported by the Research Council of Norway through its Centres of Excellence scheme, project no. 262622 and through grants of computing time from the Programme for Supercomputing.
\end{acknowledgements} 

\bibliography{waveguides}

\end{document}